\documentclass[journal,technote]{IEEEtran}

\usepackage[cmex10]{amsmath}

\usepackage{amssymb}

\newtheorem{theorem}{Theorem}
\newtheorem{lemma}[theorem]{Lemma}
\newtheorem{definition}[theorem]{Definition}
\newtheorem{remark}[theorem]{Remark}

\def\wt{\mathop{{\rm wt}}}
\def\F{{\mathbb F}}

\def\theoremcite#1#2{\cite[#1]{#2}}

\def\bm#1{\mathchoice{\mbox{\boldmath{$\displaystyle #1$}}}%
{\mbox{\boldmath{$\textstyle #1$}}}%
{\mbox{\boldmath{$\scriptstyle #1$}}}%
{\mbox{\boldmath{$\scriptscriptstyle #1$}}}}

\begin{document}
\title{The Weights in MDS Codes}

\author{Martianus~Frederic~Ezerman,~\IEEEmembership{Student Member,~IEEE,}\linebreak[3]%
        Markus~Grassl,~\IEEEmembership{Member,~IEEE,}\linebreak[3]%
        and~Patrick~Sol{\'e},~\IEEEmembership{Member,~IEEE}
\thanks{M. F. Ezerman is with the Division of Mathematical Sciences,
  School of Physical and Mathematical Sciences, Nanyang Technological
  University, Singapore 637371, Republic of Singapore (e-mail:
  mart0005@ntu.edu.sg).}%
\thanks{M. Grassl is with the Centre for Quantum Technologies,
  National University of Singapore, Singapore 117543, Republic of
  Singapore (e-mail: markus.grassl@nus.edu.sg).}%
\thanks{P. Sol{\'e} is with the Centre National de la Recherche
  Scientifique (CNRS), Laboratoire LTCI, Telecom-ParisTech, Dept Comelec, 46 rue
  Barrault, 75634 Paris, cedex France (e-mail: sole@enst.fr).}%
}

\markboth{submitted to IEEE Transactions on Information Theory, revised July 2010 }%
{Ezerman \MakeLowercase{\textit{et al.}}: The Weights in MDS Codes}

\maketitle

\begin{abstract}
The weights in MDS codes of length $n$ and dimension $k$ over the
finite field $GF(q)$ are studied.  Up to some explicit exceptional
cases, the MDS codes with parameters given by the MDS conjecture are
shown to contain all $k$ weights in the range $n-k+1$ to $n$. The
proof uses the covering radius of the dual code.
\end{abstract}


\section{Introduction}\label{sec:introduction}
One of the main properties of a linear error-correcting code is its
minimum distance since this determines the maximal number of errors
that can be corrected independently of the position of the errors.
More information about the error-correcting properties of a code can
be derived from its weight distribution.

In some cases, one does not need to know the number of codewords of a
given weight, but only the set of weights of the codewords.  Assmus
and Mattson \cite{AsMa69} established a connection between codes and
designs based on the non-zero weights in the codes.  Hill and Lizak
\cite{HiLi95,Hi99} derived conditions on the non-zero weights that
imply that a code can be extended.  Rains \cite{Rai99} showed that a
quantum error-correcting code (QECC) of length $n'<n$ can be derived
from a QECC of length $n$ if an auxiliary code contains a word of
weight $n'$.  Using this result, R\"otteler \textit{et al.}
\cite{GBR04,RGB04} constructed quantum MDS codes for all lengths $n\le
q+1$.  The construction relies on the following statement which,
\textit{e.\,g.}, can be found in \cite[p.~320]{MS77} without an
explicit proof:
\begin{equation}\label{eq:MS77}
\begin{array}[b]{p{0.8\hsize}}
\text{An MDS code with parameters $[n,k,d]_q$ has}\\
\text{$k$ distinct nonzero weights, $n-k+1,\ldots,n$.}
\end{array}
\end{equation}
In this note, we show that this statement is not true in general and
investigate for which parameters of MDS codes it holds.

The material is organized as follows. Section \ref{sec:preliminaries}
collects the necessary definitions and notation (for further details
see, \textit{e.\,g.}, \cite{HuPl03,MS77}). Section
\ref{sec:trivialMDS} disposes of the trivial codes. Section
\ref{sec:length_q} studies codes with $n \le q$. Section
\ref{sec:length_q+1} and Section \ref{sec:length_q+2} study codes of
lengths $q+1$ and $q+2$, respectively.

\section{Preliminaries}\label{sec:preliminaries}
A linear code with parameters $[n,k,d]_q$ of length $n$, dimension
$k$, and minimum distance $d$ is a subspace of dimension $k$ of the
vector space $\F_q^n$ over the finite field $\F_q=GF(q)$ with $q$
elements.  Any two vectors in $C$ differ in at least $d$ places.  From
the Singleton bound we have $d\le n-k+1$, and a code establishing this
bound is called a \emph{maximum distance separable} (MDS) code.
Trivial families of MDS codes are the full vector space $[n,n,1]_q$,
repetition codes $[n,1,n]_q$ of any length, and their duals
$[n,n-1,2]_q$.  The main conjecture on MDS codes states that if there
is a non-trivial MDS code with parameters $[n,k,n-k+1]_q$ over $\F_q$,
then $n\le q+1$, except when $q$ is even and $k=3$ or $k=q-1$ in which
case $n \le q+2$ (see, \textit{e.\,g.}, \cite[Chapter 7.4,
  p.~265]{HuPl03}).

The weight enumerator $W_C(X,Y)$ of a code is given by the polynomial
\[
W_C(X,Y)=\sum_{w=0}^n A_w X^{n-w}Y^w,
\]
where $A_w$ is the number of codewords of Hamming weight $w$ in the
code $C$. The weight enumerator of an MDS code is unique.  It is given
by $A_0=1$, $A_w=0$ for $0<w<d$, and
\begin{alignat}{3}
A_w&{}=\binom{n}{w}\sum_{j=0}^{w-d}(-1)^j \binom{w}{j}(q^{w-d+1-j}-1)\nonumber\\
&{}=\binom{n}{w}(q-1)\sum_{j=0}^{w-d}(-1)^j \binom{w-1}{j}q^{w-d-j},\label{eq:MDS_distribution}
\end{alignat}
for $d\le w\le n$ (see, \textit{e.\,g.}, \cite[Ch.~11, \S3, Theorem~6]{MS77}).

While this formula allows us to compute the weight enumerator of
any MDS code, it is not obvious for which weights $w$ there is a
codeword of weight $w$ in the code $C$, \textit{i.\,e.}, $A_w>0$.

\section{Trivial MDS Codes}\label{sec:trivialMDS}
Statement (\ref{eq:MS77}) is clearly true for the trivial MDS codes
with parameters $[n,1,n]_q$ and $[n,n,1]_q$ over any field $\F_q$.
For the dual of the repetition code, we have the first counterexample.
\begin{theorem}\label{th:trivial}
The code with parameters $[n,n-1,2]_q$ contains words of all weights
$2,\ldots,n$ if and only if $q>2$ or $n=2$.
\end{theorem}
\begin{IEEEproof}
The dual of the binary repetition code of length $n$ is an even weight
code with parameters $[n,n-1,2]_2$.  It contains only words of even
weight, so statement (\ref{eq:MS77}) is false for $q=2$ and $n>2$.

Now let $q>2$.  The codewords of the code $C$ with parameters
$[n,n-1,2]_q$ are exactly those vectors for which the sum of the
coefficients is zero. We prove by induction that $C$ contains vectors
of all weights $w=2,\ldots,n$.  For $w=2$, the code contains the
vector $\bm{v}^{(2)}=(1,-1,0,0,\ldots,0)$. Now assume that there
exists a vector $\bm{v}^{(w)}\in C$ with $\wt(\bm{v}^{(w)})=w<n$.
Without loss of generality, let the first $w$ coordinates of
$\bm{v}^{(w)}$ be non-zero. We construct a vector $\bm{v}^{(w+1)}\in
C$ with $\wt(\bm{v}^{(w+1)})=w+1$ as follows. Replace the non-zero
element $v^{(w)}_w$ by a different non-zero element of $\F_q$ to
obtain a vector $\bm{x}$ with $\sum_i x_i=s\ne 0$. Next, replace
$x_{w+1}=0$ by $-s$. The resulting vector is the desired new codeword
$\bm{v}^{(w+1)}$.
\end{IEEEproof}

\section{MDS Codes of length $n\le q$}\label{sec:length_q}
In this section we show that statement (\ref{eq:MS77}) is true for all
MDS codes of length $n\le q$.  The main tool is a relation, due to
Delsarte \cite{Del73}, between the external distance $s'$ of a code
and the covering radius which we define first.
\begin{definition}[Covering radius]
The covering radius $\rho(C)$ of a linear code with parameters
$[n,k,d]_q$ is the maximal distance of any vector in $\F_q^n$ to the
code $C$, \textit{i.\,e.}
\[
\rho(C)=\max_{\bm{v}\in\F_q^n}\:\min_{\bm{c}\in C} d(\bm{v},\bm{c}).
\]
\end{definition}
\begin{definition}[External distance]
The external distance $s'$ of a linear code with parameters $[n,k,d]_q$ is the
number of non-zero weights in the dual code $C'=C^\bot$, \textit{i.\,e.}
\[
s'=|\{i\colon i=1,\ldots,n|B_i\ne 0\}|,
\]
where $B_i$ denotes the number of codewords of weight $i$ in the dual
code $C^\bot$.  The dual code $C^\bot$ is given by
\[
C^\bot=\{\bm{v}\in\F_q^n\colon\bm{v}\cdot\bm{c}=\sum_{i=1}^n
v_ic_i=0\; \forall \bm{c}\in C\}.
\]
\end{definition}
\begin{theorem}[External distance bound \theoremcite{Theorem~3.3}{Del73}]\label{theorem:Delsarte}\ \\
For a code with parameters $[n,k,d]_q$ with external distance $s'$,
every vector in $\F_q^n $ is at distance less than or equal to $s'$
from at least one codeword. Hence, for the covering radius $\rho(C)$,
we have $\rho(C)\le s'$.
\end{theorem}

This implies that statement (\ref{eq:MS77}) holds for codes $C$ for
which the covering radius $\rho(C^\perp)$ of the dual code $C^\bot$ is
at least $k$.  Then by Theorem \ref{theorem:Delsarte}, the number $s$
of nonzero weights in $C$ is at least $s\ge k$.  Trivially, we have
$s\le n-d+1$ and therefore $s\le k$ for MDS codes.

Furthermore, we use the following ``supercode lemma'' which can,
\textit{e.\,g.}, be found in \cite[Lemma 8.2.1]{CHLL97}.
\begin{lemma}\label{lemma:supercode}
Let $C_0\subset C$ and $D(C_0,C)$ be the maximum distance of a vector
in $C$ to $C_0$. Then $\rho(C_0)\ge D(C_0,C)$. In particular,
$\rho(C_0)\ge d(C)$.
\end{lemma}

With this preparation, we are ready to prove the following.
\begin{theorem}\label{theorem:length_q}
Any MDS code with parameters $[n,k,d]_q$ of length $n\le q$ has $k$
nonzero weights.
\end{theorem}
\begin{IEEEproof}
Trivially, a code with minimum distance $n-k+1$ can have at most $k$
non-zero weights.  The extended narrow-sense Reed-Solomon (RS) codes
with parameters $[q,k,q-k+1]_q$ form a sequence of nested MDS codes
\cite[Theorem 5.3.2]{HuPl03}. Shortening these codes, we obtain
sequences of nested MDS codes $C_{n,k}=[n,k,n-k+1]_q$ with
$C_{n,k}\subset C_{n,k+1}$ for any length $n\le q$.  For the dual
codes we have $C_{n,k}^\perp=[n,n-k,k+1]_q\subset
C_{n,k-1}^\perp=[n,n-k+1,k]_q$.  By Lemma~\ref{lemma:supercode} the
code $C_{n,k}^\perp$ has covering radius $\rho(C_{n,k}^\perp)\ge
k$. Using Theorem~\ref{theorem:Delsarte}, it follows that $s\ge k$,
\textit{i.\,e.}, the code $C$ has at least $k$ non-zero weights.
Recall that the weight distribution of an MDS code with parameters
$[n,k,d]_q$ is uniquely determined by its parameters. Hence the result
does not only hold for the MDS codes derived from extended RS codes,
but for all MDS codes.
\end{IEEEproof}
\begin{remark}
The covering radius of extended RS codes is also given in
\cite[Theorem 10.5.7]{CHLL97}.  The lower bound on the covering radius
of MDS codes of length $n\le q$ can also be found in \cite{GaKl98}.
\end{remark}

\section{Codes of Length $q+1$}\label{sec:length_q+1}
In \cite[Theorem~2]{GaKl98} it has been shown that the covering radius
of some MDS codes with parameters $[q+1,k,d]_q$ is $\rho(C)=d-2$.  In
this case, Theorem~\ref{theorem:Delsarte} only implies that there are
at least $k-1$ non-zero weights, and there is indeed a family of MDS
codes of length $q+1$ for which statement (\ref{eq:MS77}) is false,
namely the simplex codes with parameters $[q+1,2,q]_q$ over $\F_q$
which contain only words of weight zero or $q$ (see, \textit{e.\,g.},
\cite[Theorem 2.7.5]{HuPl03}).

Furthermore, if $\rho(C)=d-2$, we can not have a sequence of nested MDS
codes of length $q+1$ with co-dimension one.  However, there are
sequences of nested MDS codes with co-dimension two.  For this, we
recall the explicit construction of MDS codes of length $q+1$ as
cyclic or consta-cyclic codes (see also \cite{GrGu08,RGB04}).
\begin{theorem}\label{theorem:cyclicMDS}
For any $k$, $1\le k\le q+1$, there exists an MDS code $C_{q+1,k}$
over $\F_q$ with parameters $[q+1,k,q-k+2]_q$ that is either cyclic or
consta-cyclic.  The codes of even dimension and the codes of odd
dimension form two sequences of nested codes, \textit{i.\,e.},
$C_{q+1,k}\subset C_{q+1,k+2}$.
\end{theorem}
\begin{IEEEproof}
Let $\omega$ denote a primitive element of $\F_{q^2}$. Hence
$\alpha:=\omega^{q-1}$ is a primitive $(q+1)$-th root of unity.

First we consider the case when $q+1-k$ is odd. We define the
following polynomial of degree $2\mu+1$:
\[
g_1(z):=\prod_{i=-\mu}^\mu (z-\alpha^i).
\]
Its zeros $\alpha^i$ and $\alpha^{-i}$ are conjugates of each other
since $\alpha^q=\alpha^{-1}$. Hence, $g_1(z)$ is a polynomial over
$\F_q$. The resulting cyclic code $C$ over $\F_q$ has length $q+1$ and
dimension $q-2\mu$. The generator polynomial $g_1(z)$ has $2\mu+1$
consecutive zeros, so the BCH bound yields $d\ge 2\mu+2$. Therefore
$C$ is an MDS code $[q+1,q-2\mu,2\mu+2]_q$.

If $q+1-k$ is even and $q$ is even too, the polynomial
\[
g_2(z):=\prod_{i=q/2-\mu}^{q/2+1+\mu} (z-\alpha^i)
=\prod_{i=q/2-\mu}^{q/2} (z-\alpha^i)(z-\alpha^{-i})
\]
has degree $2\mu+2$. It is a polynomial over $\F_q$ with $2\mu+2$
consecutive zeros, so the resulting code is an MDS code with parameters
$[q+1,q-1-2\mu,2\mu+3]_q$.

Finally, if $q+1-k$ is even and $q$ is odd, consider the polynomial
\[
g_3(z):=\prod_{i=1}^\mu(z-\omega \alpha^i)(z-\omega \alpha^{1-i})
\]
of degree $2\mu$. The roots $\omega \alpha^i$ and $\omega
\alpha^{1-i}$ are conjugates of each other as
$(\omega\alpha^i)^q=\omega^{(1+(q-1)i)q}
=\omega^{q+(1-q)i}=\omega^{1+(q-1)(1-i)}=\omega\alpha^{1-i}$, so
$g_3(z)$ is a polynomial over $\F_q$. Furthermore, $g_3(z)$ divides
$z^{q+1}-\omega^{q+1}\in\F_{q^2}[z]$ as
$(\omega\alpha^i)^{q+1}=\omega^{q+1}$.  Therefore $g_3(z)$ defines a
consta-cyclic code $C$ of length $q+1$ and dimension $q+1-2\mu$ over
$\F_q$. Note that $C$ can be considered as a shortened subcode of the
cyclic code of length $q^2-1$ over $\F_{q^2}$ generated by
$g_3(z)$. For the latter, the BCH bound yields $d\ge 2\mu+1$, hence
$C$ is an MDS code with parameters $[q+1,q+1-2\mu,2\mu+1]_q$.

The statement $C_{q+1,k}\subset C_{q+1,k+2}$ follows from the
particular form of the polynomials $g_i(z)$.
\end{IEEEproof}
\begin{remark}
Theorem \ref{theorem:cyclicMDS} is a slightly modified version of
Theorem 9 in \cite[Ch.~11, \S5]{MS77}. There only cyclic codes are
considered; the construction fails when both $q$ and $k$ are odd (see
also the preface to the third printing of \cite{MS77}).
\end{remark}

\begin{theorem}\label{theorem:n=q+1}
An MDS code with parameters $[q+1,k,d]_q$ has $k$ nonzero weights,
except when $k=2$.
\end{theorem}
\begin{IEEEproof}
The statement is clearly true for $k=1$, and it does not hold for
$k=2$ since the code with parameters $[q+1,2,q]_q$ is a simplex code.
For $k>2$, let $C^{(p)}=[q,k,d-1]_q$ and $C^{(s)}=[q,k-1,d]_q$ be the
MDS codes obtained by puncturing and shortening of $C$, respectively.
Without loss of generality, we may assume that we have deleted the
last position.  By Theorem~\ref{theorem:length_q}, $C^{(s)}$ contains
codewords of all $k-1$ weights $q-k+2,\ldots,q$.  Appending zero to
the codewords in $C^{(s)}$, we obtain a subcode of $C$ with the same
weight distribution as that of $C^{(s)}$.  Thus it remains to show
that $C$ contains a word of weight $q+1$.  If the dimension of $C$ is
odd, by Theorem~\ref{theorem:cyclicMDS} the code $C$ contains a
subcode $[q+1,1,q+1]_q$ and hence a word of weight $q+1$.  If the
dimension of $C$ is even, we know that statement (\ref{eq:MS77}) is
false for dimension $k=2$.  If the dimension is at least four, the
code $C$ contains a subcode $[q+1,4,q-2]_q$.  For this code of
dimension four, using (\ref{eq:MDS_distribution}) we compute
\[
A_{q+1}=(q-1)\sum_{j=0}^3(-1)^j \binom{q}{j}q^{3-j}=\frac{1}{3}q(q-1)(q^2-1).
\]
This shows that the code with parameters $[q+1,4,q-3]_q$ contains
words of weight $q+1$.
\end{IEEEproof}

\section{Codes of Length $q+2$}\label{sec:length_q+2}
MDS codes of length $q+2$ are known for $q=2^m$, and $k=3$ or
$k=2^m-1$ (see, \textit{e.\,g.}, \cite[Theorem 10, Ch.~11,
  \S5]{MS77}). For $m>1$, a generator matrix or parity check matrix is
given by
\begin{equation}\label{eq:paritycheck}
H=\left(
\begin{array}{*{12}{l}}
1 & 1        & 1        & 1       &\dots & 1             &  1 & 0 & 0\\
1 & \alpha   & \alpha^2 & \alpha^4 &\dots & \alpha^{q-2}   & 0 & 1 & 0\\
1 & \alpha^2 & \alpha^4 & \alpha^6 &\dots & \alpha^{2(q-2)} & 0 & 0 & 1
\end{array}
\right),
\end{equation}
where $\alpha$ is a primitive element of the field $GF(2^m)$.

First consider the code with parameters $[2^m+2,3,2^m]_{2^m}$.  Assume
that there is a codeword $\bm{v}\in C$ of weight $n-1=2^m+1$.
Shortening $C$ at the position where $\bm{v}$ is zero, we get a code
$C^{(s)}=[2^m+1,2,2^m]_{2^m}$ which contains a codeword $\bm{v}'$ of
weight $2^m+1$. But we already know that the code $C^{(s)}$ is a
$q$-ary simplex code which does not contain a codeword of weight
$2^m+1$.  Hence the code with parameters $[2^m+2,3,2^m]_{2^m}$ does
not contain a word of weight $2^m+1$.  Using
(\ref{eq:MDS_distribution}), the non-zero coefficients of the weight
enumerator are computed as $A_{2^m}=(2^{2m}-1)(2^{m-1}+1)$,
$A_{2^m+2}=2^{m-1}(2^m-1)^2$, and $A_0=1$.

Next, consider the dual code with parameters $[2^m+2,2^m-1,4]_{2^m}$
with the parity check matrix $H$ given in (\ref{eq:paritycheck}).
\begin{theorem}
The MDS code with parameters $[2^m+2,2^m-1,4]_{2^m}$ contains words of all weights
$w=4,\ldots,2^m+2$ if and only if $m\ne 2$.
\end{theorem}
\begin{IEEEproof}
For $m=1$, the code is the binary repetition code $[4,1,4]_2$ for
which the statement holds.  For $m=2$, we obtain the hexacode, a
famous two-weight code (see \cite[Chapter 3, (2.5.2)]{CoSl98}).  So
let $m>2$.  Similar to the proof of Theorem~\ref{theorem:n=q+1},
considering the shortened code $C^{(s)}=[2^m+1,2^m-2,4]_{2^m}$ we find
that the code $C$ contains words of all weights $w=4,\ldots,2^m+1$. It
remains to show that $C$ contains a word $\bm{v}$ of weight $2^m+2$.
Consider the following vector
\[
\bm{v}=(\alpha,\underbrace{1,\ldots,1}_{q-2},\alpha,\alpha+1,\alpha+1),
\]
where $q=2^m$.  In order to show that $\bm{v}$ is in the kernel of
$H$ and hence $\bm{v}\in C$, we note that $\sum_{i=0}^{q-2}
(\alpha^j)^i=0$ for $0<j<q-1$.
\end{IEEEproof}

\section{Conclusions}
In summary, we have the following:
\begin{itemize}
\item The trivial MDS codes with parameters $[n,n,1]_q$, $[n,1,n]_q$,
  and $[n,n-1,2]_q$ have $k$ non-zero weights with the exception of
  the dual of the binary repetition code of length $n>2$ which
  contains only words of even weights.
\item The MDS codes with parameters $[n,k,d]_q$ of length $n\le q+1$
  have $k$ non-zero weights, with the exception of the $q$-ary simplex
  code with parameters $[q+1,2,q]_q$ which contains only words of
  weight zero or $q$.
\item For $m\ne 2$, the codes with parameters $[2^m+2,2^m-1,4]_{2^m}$
  have $k=2^m-1$ non-zero weights.  These codes are quasi perfect with
  covering radius $2$.
\item The code with parameters $[2^m+2,3,2^m]_{2^m}$ has only non-zero
  codewords of weight $2^m$ and $2^m+2$, with
  $A_{2^m}=(2^{2m}-1)(2^{m-1}+1)$ and $A_{2^m+2}=2^{m-1}(2^m-1)^2$.
  These two-weight codes are known as family TF1 in \cite{CaKa86}.
\end{itemize}

Our result covers the parameters of all non-trivial MDS codes given by
the MDS conjecture.  In general, if a non-trivial MDS code with
parameters $[n,k,d]_q$ exists, then $2\le k \le \min\{n-2,q-1\}$ and
$n\le q+k-1\le 2q-2$ (see \cite[Corollary~7.4.4]{HuPl03}).

Finally, we note that our result confirms the construction of quantum
MDS codes given in \cite{GBR04,RGB04} as statement (\ref{eq:MS77})
holds for all MDS codes used therein to derive shortened quantum
codes.
\section*{Appendix: An Alternative Approach}
After the preliminary draft of this paper was made public, we have 
received useful comments and suggestions, some of which are 
instructive to better understand the weights in MDS codes.

Ludo Tolhuizen showed in 1988 in an unpublished research
report that $A_w\ge f(q,n,d,w)$ where
\begin{equation*}
f(q,n,d,w)={n \choose w}(q-d)(q-1)^{w-d} > 0
\end{equation*}
for $d<q$.

A referee observed that most results shown in this paper can also be 
derived through a careful analysis of Equation (\ref{eq:MDS_distribution}). 
Let us rewrite the said equation as
\begin{equation}\label{eq:MDS_Alt}
A_w=\binom{n}{w}(q-1)\sum_{j=0}^{w-d}(-1)^j a_{j} 
\text{, with } a_{j}=\binom{w-1}{j}q^{w-d-j}
\end{equation}
for $d\le w\le n$ and investigate the cases for which $A_{w}=0$.

First we show that for the $[n,n-1,2]_{2}$-code $C$ with 
$n>2$, $A_{w}=0$ if and only if $w$ is odd. Note that if 
$l$ is an odd positive integer and
\begin{equation*}
S:=\sum^{l}_{j=0}(-1)^{j} \binom{l+1}{j}2^{l-j} \text{,}
\end{equation*}
we have
\begin{equation*}
 2S+1=\sum^{l+1}_{j=0}(-1)^{j} \binom{l+1}{j}2^{l+1-j}=(2-1)^{l+1}=1 \text{,}
\end{equation*}
which implies that $S=0$. If $l$ is an even positive integer, 
it follows that
\begin{equation*}
S: =\sum^{l}_{j=0}(-1)^{j} \binom{l+1}{j}2^{l-j} = l+1 \text{.}
\end{equation*}
Thus, we have an alternative proof for the assertion regarding the 
dual of the binary repetition code in Theorem~\ref{th:trivial}.

For the nontrivial cases, observe that
\begin{equation*}
 R_{j}:=\frac{a_{j}}{a_{j+1}}= q \frac{j+1}{w-1-j}\text{.}
\end{equation*}
Clearly, for $w\leq q$ we have
$R_{j}>1,$ which implies that $A_w>0$ for $n\leq q$.

For $n=q+1$, $R_{j} = 1$ if and only if $j=0$ and $w=q+1$ 
since $q (j+1) > w-1-j$ for $j \geq 1$. Hence, $d=q$ and $k=2$.
Therefore, an MDS code with parameters $[q+1,2,q]_{q}$ has $A_{q+1}=0$.

Similarly for $n=q+2$ with $q=2^{m}$. $R_{j} = 1$ if and only if $j=0$ 
and $w=q+1$, forcing $d=q$ and $k=3$. There are no codewords of 
weight $2^{m}+1$ in any MDS code with parameters 
$[2^{m}+2,3,2^{m}]_{2^{m}}$.

\section*{Acknowledgment}
The authors thank Fr{\'e}d{\'e}rique Oggier for hosting
the discussion that lead to this paper in her office on August~5,
2009, and Roxana Smarandache for helpful discussions. We also thank 
the referees for their insightful suggestions.

Patrick Sol\'e is funded in part by the Singapore National
Research Foundation under Research Grant NRF-CRP2-2007-03. Centre 
for Quantum Technologies is a Research Centre of Excellence
funded by the Ministry of Education and the National Research 
Foundation of Singapore.

\begin{IEEEbiographynophoto}{Martianus~Frederic~Ezerman}
received his BA in Philosophy and BSc in Mathematics in 2005 and his
MSc in Mathematics in 2007, all from Ateneo de Manila University,
Philippines. He is currently a PhD candidate under research
scholarship at the Division of Mathematical Sciences, School of
Physical and Mathematical Sciences, Nanyang Technological University,
Singapore. His research interests include coding theory, focusing on
quantum error-correcting codes, and low PMPR
(peak-to-mean-power-ratio) sequences.
\end{IEEEbiographynophoto}

\begin{IEEEbiographynophoto}{Markus Grassl}
received his diploma degree in Computer Science in 1994 and his
doctoral degree in 2001, both from the Fakult\"at f\"ur Informatik,
Universit\"at Karlsruhe (TH), Germany.  His dissertation was on
constructive and algorithmic aspects of quantum error-correcting
codes.

From 1994 to 2007 he was a member of the Institut f\"ur Algorithmen
und Kognitive Systeme, Fakult\"at f\"ur Informatik, Universit\"at
Karlsruhe (TH), Germany.  From 2007 to 2008 he was with the Institute
for Quantum Optics and Quantum Information of the Austrian Academy of
Sciences in Innsbruck.  In 2009, he joined the Centre for Quantum
Technologies at the National University of Singapore.

His research interests include quantum computation, focusing on
quantum error-correcting codes, and methods of computer algebra in
algebraic coding theory.  He maintains tables of good block quantum
error-correcting codes as well as good linear block codes.
\end{IEEEbiographynophoto}

\begin{IEEEbiographynophoto}{Patrick Sol{\'e}}
received the Ing\'enieur and Docteur-Ing{\'e}nieur degrees both from
Telecom ParisTech, Paris, France, in 1984 and 1987, respectively, and
the habilitation \`a diriger des recherches from Universit\'e de
Nice-Sophia Antipolis, Sophia Antipolis, France, in 1993.

He has held visiting positions in Syracuse University, Syracuse, NY,
from 1987 to 1989, Macquarie University, Sydney, Australia, from 1994
to 1996, and Lille University, Lille, France, from 1999 to 2000. From 
1989 to 2009, he has been a permanent member of the CNRS
Laboratory I3S, Sophia Antipolis, France, and from 2009 to present of
CNRS Laboratory LTCI, Paris, France.

His research interests include coding theory (covering radius, codes
over rings, geometric codes), interconnection networks (graph spectra,
expanders), vector quantization (lattices), and cryptography (Boolean
functions). Dr. Sol{\'e} is the recipient (jointly with Hammons, Kumar,
Calderbank, and Sloane) of the IEEE Information Theory Society Best
Paper Award in 1994. He has served as an associate editor of the
Transactions from 1999 till 2003.
\end{IEEEbiographynophoto}


\end{document}